\documentclass[epj]{svjour}
\newcommand{\be}{\begin{equation}}
\newcommand{\ee}{\end{equation}}
\newcommand{\bea}{\begin{eqnarray}}
\newcommand{\eea}{\end{eqnarray}}
\usepackage{graphicx,amssymb,amsmath,epsfig,bm,color,slashed,mathrsfs}
\usepackage{amsmath,txfonts,multirow}
\usepackage[square,sort,comma,numbers]{natbib}

\def\apj{ApJ}%
\def\apjl{ApJ}%
\def\aap{A\&A}%
\def\mnras{MNRAS}%
\def\prc{Phys.~Rev.~C}%
\def\prd{Phys.~Rev.~D}%
\def\nat{Nature}%
\begin{document}
\title{Cooling compact stars and phase transitions in dense QCD}
\author{Armen Sedrakian
}                     
\offprints{sedrakian@th.physik.uni-frankfurt.de}          
\institute{\it Institute for Theoretical  Physics,
 J. W. Goethe University,\\
Max-von-Laue-Str. 1,
D-60438  Frankfurt am Main, Germany
}
\date{Received: date / Revised version: date}
%
\abstract{
 We report new simulations of cooling of compact stars
  containing quark cores and updated fits to the Cas A fast cooling
  data. Our model is built on the assumption that the transient
  behaviour of the star in Cas A is due to a phase transition within
  the dense QCD matter in the core of the star.  Specifically, the
  fast cooling is attributed to an enhancement in the neutrino
  emission triggered by a transition from a fully gapped, two-flavor,
  red-green color-superconducting quark condensate to a
  superconducting crystalline or an alternative gapless,
  color-superconducting phase. The blue colored condensate is modeled
  as a Bardeen-Cooper-Schrieffer (BCS)-type color superconductor with
  spin-one pairing order parameter. We study the sensitivity of the
  fits to the phase transition temperature, the pairing gap of blue
  quarks and the time-scale characterizing the phase transition (the
  latter modelled in terms of a width parameter).  Relative variations
  in these parameter around their best fit values larger than
  $10^{-3}$ spoil the fit to the data.  We confirm the previous
  finding that the cooling curves show significant variations as a
  function of compact star mass, which allows one to account for
  dispersion in the data on the surface temperatures of thermally
  emitting neutron stars.
} 
\maketitle

\section{Introduction}
\label{intro}
The central densities of compact stars (hereafter CSs) can be by a factor of few up to
ten times larger than the saturation density of nuclear matter. In
this range of densities the mean interparticle distance may become of
the order of characteristic size of a baryon, therefore neutrons,
protons, and heavier baryons will eventually lose their identity, as
their wave-functions start to overlap.  Although the detailes of the
mechanism of deconfinement are not well understood so far, it is plain
that at sufficiently large density the baryonic matter will undergo a
deconfinement phase transition to quark matter. Not only the
transition itself, but also the phase structure of quark matter is
difficult to access, because the densities in CSs  will not
be large enough for the perturbative QCD to be valid. Astrophysics of
CSs  offers an avenue to explore and constrain the possible
deconfinement phase transition and the properties of dense QCD matter
via modeling of CSs  containing quark
cores~\cite{2014arXiv1404.3723B,2014JPhG...41l3001B}.

The focus of this work is the effects of the deconfinement phase
transition and the phase structure of dense QCD on the cooling of
CSs. This discussion extends the previous work on the
cooling of hybrid stars containing superconducting quark matter,
including the interpretation of the rapid cooling of the CS
in Cassiopea A as a phase transition within the QCD phase
diagram~\cite{2013A&A...555L..10S,2011PhRvD..84f3015H}.  Specifically,
we report the results of new simulations of cooling of CSs 
and fits to the Cas A updated data~\cite{2013ApJ...777...22E}, which
covers the 10 year period from 2003 to 2013. These data indicates an
unprecedented fast cooling of the neutron star in Cas A, which
requires fast transient modeling of this object (as opposed to the
familiar long-time-scale modeling of neutron stars, where characteristic
timescales of variations are $t\ge 100$~y). The best estimate of
Ref.~\cite{2013ApJ...777...22E} indicates a decline in the temperature
of the star $2.9\pm 0.9\%$ over 10 years of observation, which
requires extremely fast transient cooling of the star over this
period.  The currently available data cannot be interpreted
unambiguously; one reason is the bright and varying supernova remnant
background, which makes a definitive interpretation
difficult~\cite{2013ApJ...777...22E}. Another uncertainty arises from
adopted constraints on the temperature fitting parameters and the
uncertainties of the effective area calibration; for example, the
apparent decline in the temperature can be compensated if one allows
for variations in the emitting region size~\cite{2013ApJ...779..186P}.
In the following we leave aside the possible uncertainties in the
interpretation of the data and focus on its theoretical modeling.

A variety of models account theoretically for the Cas A data. Each
class of models has its own specific trigger for onset of rapid
cooling around the age of the Cas A ($t \simeq 300$ y).  A class of
nucleonic models (i.e., models of CSs containing neutrons, protons,
and electrons) attribute the rapid cooling to the onset of Cooper
pair-breaking process in CS's superfluid
component~\cite{2011PhRvL.106h1101P,2011MNRAS.412L.108S,2015MNRAS.446.3621S,2015PhRvC..91a5806H,2013ApJ...779L...4N,2014arXiv1411.6833L}.
The rapid cooling occurs once the temperature drops below the critical
temperature $T_{c,n}\simeq 10^9$~K of phase transition of nucleons to
the superluid state. This allows for additional neutrino emission via
Cooper-pair breaking and formation (PBF) processes from $S$-wave
~\cite{1976ApJ...205..541F,2006PhLB..638..114L,2007PhRvC..76e5805S,2008PhRvC..77f5808K,2012PhRvC..86b5803S}
and $P$-wave
condensates~\cite{1999A&A...343..650Y,2011PhRvC..84d5501L}.  These
minimalistic models do not require any additional new physics beyond
the standard scenario which is supplemented by pair-breaking
processes~\cite{2009ApJ...707.1131P}.

The Cas A data was also explained in
Ref.~\cite{2012PhRvC..85b2802B,2013addition} within a model where
modified Urca and bremsstrahlung processes are enhanced compared to
rates used in the minimalistic models quoted above by several orders
of magnitude due to a softening of pionic modes, as initially
discussed in Ref. ~\cite{1997A&A...321..591S}.  Rotational changes in
star's composition may induced Urca process as the star slows down,
which in turn can cause a Cas A type fast
transient~\cite{2013PhLB..718.1176N}, which however would require very
fine tuning of parameters, as already discussed in
Ref. \cite{2013A&A...555L..10S}.

Color-superconductivity was included in the simulations of CS to
account for Cas A data in
Refs.~\cite{2013A&A...555L..10S,2013ApJ...765....1N}. In
Ref.~\cite{2013ApJ...765....1N} the phase transition from purely
nucleonic to purely quark matter occurs via mixed phases, whereas
Ref.~\cite{2013A&A...555L..10S} assumes sharp interface between the
two; which scenario is realized depends on the poorly known surface
energy between the nucleonic and quark phases. We explore here the
second possibility.

This paper is organized as follows.  In sect.~\ref{sec:PhaseDiagram} we
discuss the relevant features of the phase diagram of dense and cold
QCD. Section~\ref{sec:PhysicalInput} describes the physical input
needed for cooling simulations, including neutrino
emissivities of various phases and specific heats.  The results of
numerical simulations are presented in sect. ~\ref{sec:NumRes} together
with the fits to the Cas A data.  A summary and conclusions are
given in sect.~\ref{sec:Conclusion}.

\section{The phase structure of dense QCD and massive 
compact stars}
\label{sec:PhaseDiagram}

This section provides the background information on the QCD phase
diagram, which is necessary for the understanding of the model of
cooling of CSs presented in the following sections.  Readers
interested in the astrophysical aspects of the modeling can proceed to
sect.~\ref{sec:PhysicalInput}; those interested in the results can go
directly to sect.~\ref{sec:NumRes}.

We consider a class of hybrid CSs , which contain dense quark
cores surrounded by a nucleonic envelope, with a sharp
phase-transition interface between these phases. Stars with such
hybrid structure naturally correspond to the most massive members of
the sequence of stellar equilibria modelled with an equation of state
which contains a phase transition from nucleonic to quark
matter. These massive members must be heavy enough to account for the
recently observed two-solar mass pulsars PSR J1614-2230 and PSR
J0348+0432~\cite{2010Natur.467.1081D,2013Sci...340..448A}.

If quarks in the deconfined phase of the star are unpaired (i.e. do
not form Cooper pairs) they will cool the star by neutrino emission
rapidly to temperatures well below those observed in Cas~A
\cite{2011PhRvD..84f3015H,2000PhRvL..85.2048P,2005PhRvD..71k4011A,2006PhRvD..74g4005A,2005PhRvC..71d5801G}.
However, cold quark matter is expected to be in one of the conjectured
superconducting phases due to the attractive component in the gluon
exchange quark-quark interaction~\cite{2000hep.ph...11333R}.  The
$\beta$-equilibrium and strange quark mass shift the Fermi surfaces of
up and down quarks apart. As a consequence some non-BCS phases can now
emerge. Such non-BCS phases include the gapless two-flavor
phases~\cite{2003PhLB..564..205S,2003PhRvD..67h5024M} or, for example,
the crystalline color-super\-conducting phase
\cite{2001PhRvD..63g4016A,2014RvMP...86..509A}.  The
crystal\-line-color-super\-conduc\-tivity itself has a multitude of
realizations, which differ by the way the translational symmetry is
broken by the condensate of Cooper pairs carrying finite momentum.
For concrete calculations we will assume below the so-called
Fulde-Ferrell phase (hereafter FF phase), which is simple to model,
but is general enough to preserve a key feature of the crystalline
phases, which is the existence of gapless modes on the Fermi surfaces
of up and down
quarks~\cite{2001PhRvD..63g4016A,2014RvMP...86..509A,2009PhRvD..80g4022S,2010PhRvD..82d5029H}.

\begin{figure}[t]
\begin{center}
\includegraphics[width=8cm,height=7cm]{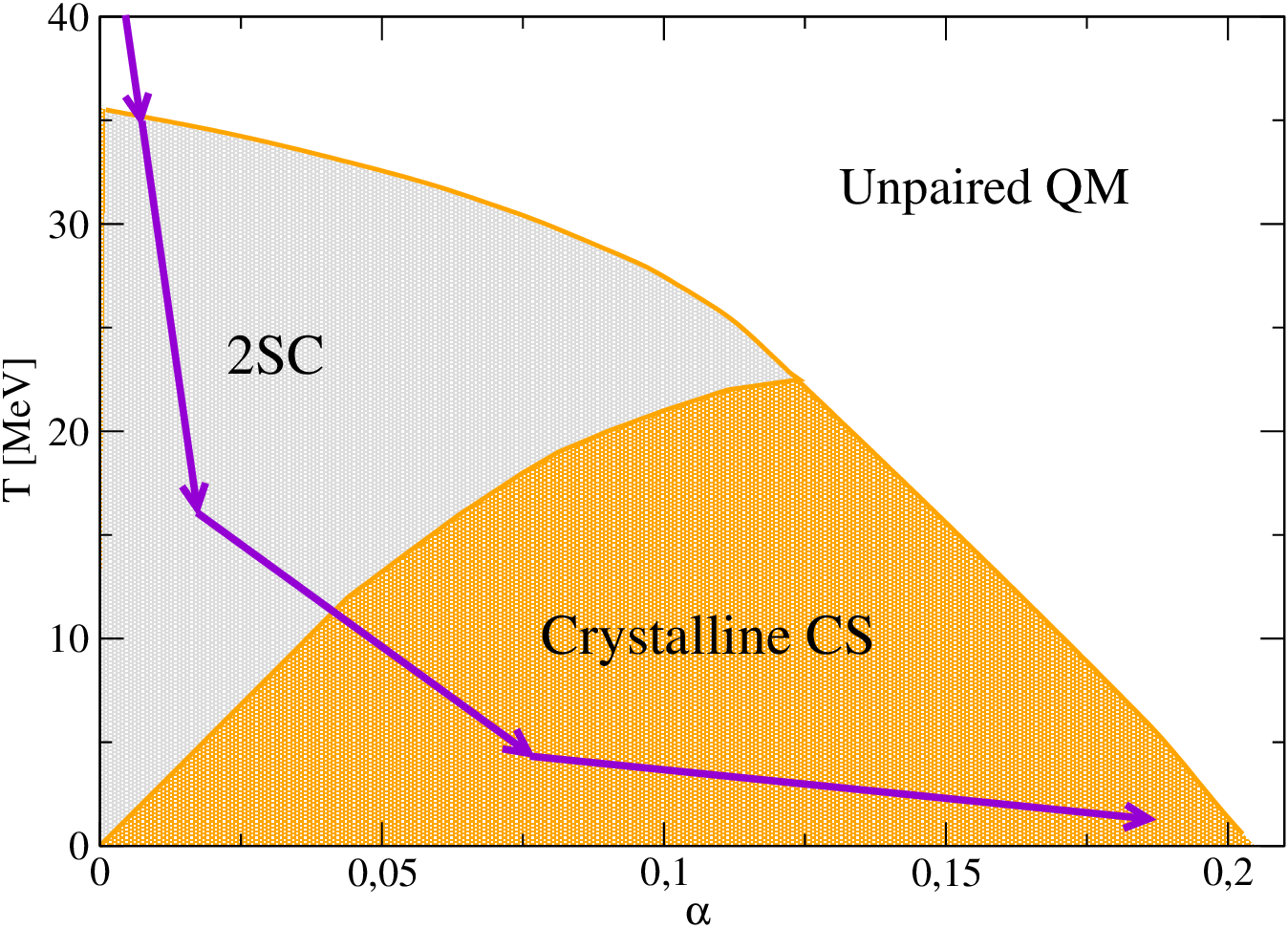}
\caption{ Phase diagram of color-superconducting 
  quark matter in the temperature ($T$) and isospin imbalance $\alpha 
  = (n_d-n_u)/(n_u+n_d)$ plane, where $n_d$ and $n_u$ are the number 
  densities of $d$ and $u$ flavors of quarks. The arrows 
  schematically show  the path in the phase diagram during the cooling of a 
  star. 
}
\label{epja:fig1}
\end{center}
\end{figure}
Our conjecture concerning the behaviour observed in Cas A is based on
a robust feature of the phase diagram of the two-flavor color
superconducting matter, (and, in fact, any two-component fermionic
systems with attractive interaction). Just below the critical
temperature and at not too high isospin asymmetries, measured by
parameter $\alpha = (n_d-n_u)/(n_u+n_d)$, where $n_d$ and $n_u$ are
the number densities of $d$ and $u$ flavors of quarks, the
rotational/translational symmetries are unbroken, i.e., the fully
gapped 2SC phase is favored (see~Fig.~\ref{epja:fig1}). At lower
temperatures a phase transition to a less symmetrical phase (such as
the FF phase) becomes favorable at larger asymmetries, but not large
enough to destroy the superconductivity completely. (This would
corres\-pond in~Fig.~\ref{epja:fig1} to $\alpha> 0.2$). Note that the
phase diagram shown in Fig.\ref{epja:fig1} is valid only in the weak
coupling limit, which is the case in the bulk of CSs . (If the coupling
is large enough the ensemble makes a transition from 2SC to
Bose-Einstein condensed phase without entering crystalline phase; see,
for example, 
Refs.~\cite{2014PhRvC..90f5804S,2014JPhCS.496a2008S,2012PhRvC..86f2801S}
for a generic and most complete phase diagram of a fermionic system with
imbalance).

\begin{figure}[tbh]
\begin{center}
\includegraphics[width=8cm,height=7cm]{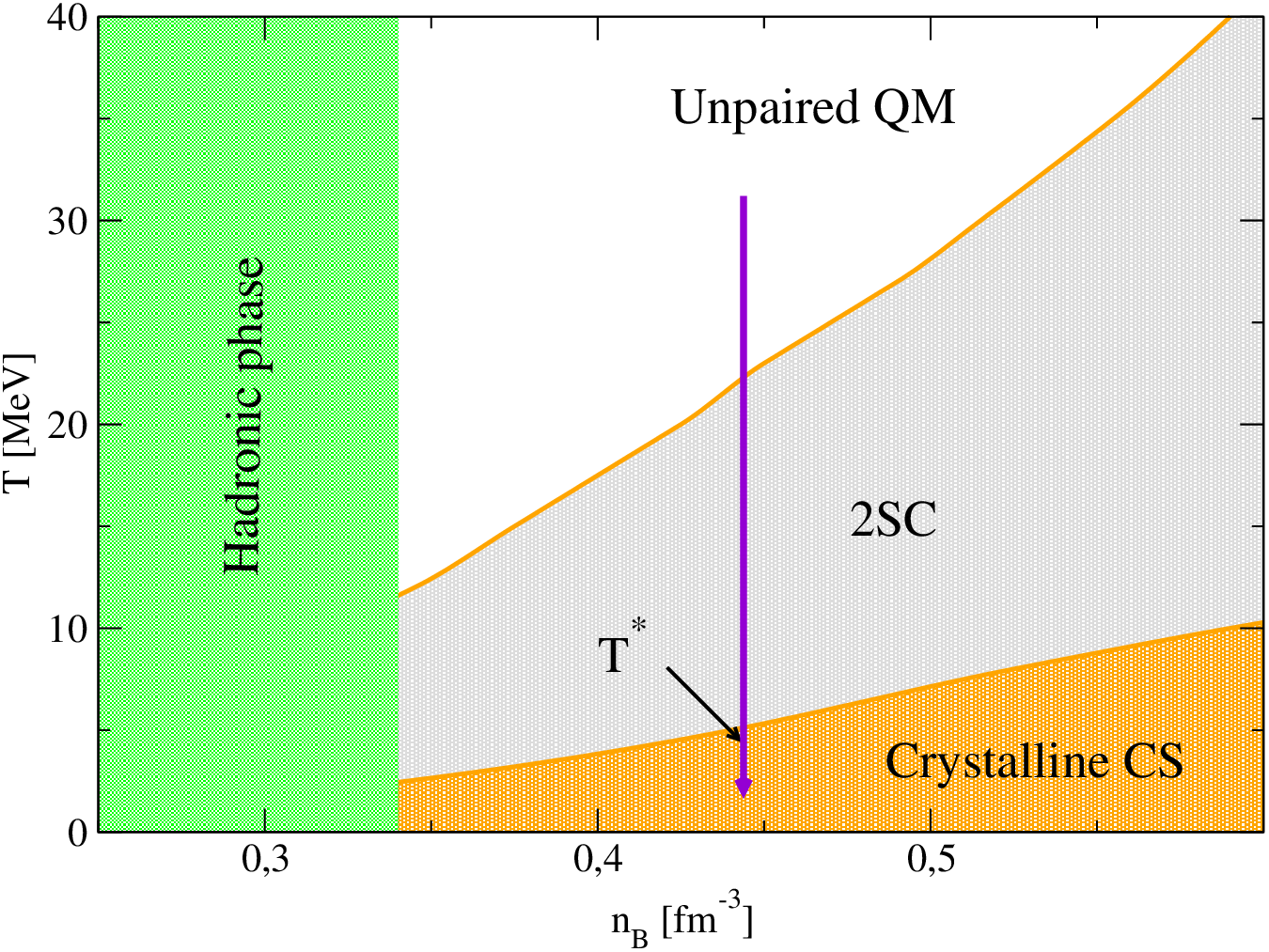}
\caption{ Phase diagram of the same phases as in fig.~\ref{epja:fig1},
  but in the temperature and baryon density ($n_B$) plane for quark
  matter in $\beta$-equilibrium with electrons. The hadronic phase is
  shown schematically on the left. }
\label{epja:fig2}
\end{center}
\end{figure}

We next explore the consequences of the transition from the
symmetrical two-flavor BCS phase to the crystalline phase (in our case
the FF phase) for the cooling of CSs and the CS in Cas A. We
anticipate such a transition in CSs because a newly born proto-CSs have 
temperatures of the order of several tens of MeV and matter in  
nearly isospin symmetrical ($\alpha \simeq 0$) state. Within short
period of time (minutes to hours) the CS cools to temperatures of
the order of 1 MeV and becomes strongly isospin asymmetrical as
$\beta$-equilibrium among quarks and electrons is established.  The
arrangement of phases shown in fig.~\ref{epja:fig1} in the case of
$\beta$-equilibrated matter is shown in fig.~\ref{epja:fig2}, which is 
adapted from Ref.~\cite{2010PhRvD..82d5029H}.
A shell of quark matter phase in the star with fixed density will
traverse the $T$-$\alpha$ plane from upper left corner to the lower
right corner as shown by arrows in fig.~\ref{epja:fig1}.  The same
path is shown in fig.~\ref{epja:fig2} in the temperature and density
plane.

 The transition temperature from the 2SC to the FF phase is
denoted by $T^*$ and will be treated below as a free parameter. Our
fits to Cas A data below will demonstrate that the values obtained from
theoretical models are larger than those obtained from fits. This
uncertainty is acceptable as the magnitude of the pairing interaction
among quarks in the non-perturbative regime of QCD is model dependent
and generally not well-know. 

Finally, assuming that the cross-flavor pairing occurs between the
green and red colored quarks we need to address the pairing among the
remaining blue colored quarks.  The strength and the flavor content of
pairing among blue quarks is model
dependent~\cite{2003PhRvD..67e4018A,2005PhRvD..71e4016S,
  2005PhRvD..72c4008A}.  Obviously blue-up and blue-down pairing is
suppressed by a large mismatch between the Fermi surfaces of the up
and down quarks.  Pairing is more likely in the color ${\bm 6}_S$ and
flavor ${\bm 3}_S$ channel, which is same-flavor and same-color
pairing, therefore it is not affected by the flavor
asymmetry~\cite{2003PhRvD..67e4018A}.  Calculations show that the gaps
are in the range 10-100 keV. Again because of uncertainties envolved
in the interaction generating the BCS state of blue quarks we treat
the blue quark gap $\Delta_b$ as a parameter of our model.


\section{Physical input into simulations}
\label{sec:PhysicalInput}

\subsection{Equilibrium structure of hybrid stars}
\label{sec:Equilibrium}

Prior to computing the time evolution of CS models with color
superconducting phases discussed in the previous section we need to
construct the equilibrium models of such stars. Two classes of models
were developed and used by us in the studies of color superconducting
quark matter. Both classes are based on equations of state (hereafter
EoS) with a sharp phase transition between the nuclear core and quark
phases; the first class features crystalline color superconducting
quark matter in the quark core and nuclear envelope, which is
described by a relativistic density functional model with a stiff
parameterization of the
EoS~\cite{2008PhRvD..77b3004I,2009PhRvD..79h3007K}. The second class
of models contains 2SC phase discussed in the previous section and the
color flavor locked (CFL) phase at high
densities~\cite{2012A&A...539A..16B,2013A&A...559A.118A}. Both models
produce stable sequences of color superconducting CSs with maximal
masses above the lower observational bound mentioned above.  We model
the high-density matter EoS within the effective Nambu-Jona- Lasinio
four-fermion interaction model supplemented by pairing interaction
between quarks.  The nuclear and quark equations of state were matched
by a Maxwell construction. If one uses alternative constructions mixed
phases will appear, which we do not discuss here
(see~Refs.~\cite{2013ApJ...765....1N,2015arXiv150706067S}).  We will
assume that strange quarks because of their large mass are not
important at the densities relevant to our models. The nucleonic phase
itself can contain non-zero strangeness in form of hyperons. But the
compatibility of hyperons with large masses of pulsars needs special
attention. Indeed their appearance softens the EoS and the maximal
masses of neutron stars are lower than those observed. This
``hyperonization puzzle'' is a fundamental problem on its own right
and will not be discussed here (see
Refs.~\cite{2013PhRvC..87e5806C,2014JPhCS.496a2003C,2014PhLB..734..383V,2015PhLB..748..369M,2015A&A...576A..68F,2015JPhG...42g5202O,2015PhRvC..92b5803B}
for recent discussions of this problem).  Below we will exclude the
hyperons from the consideration.
\begin{table}[t]
  \caption{
    The table lists the central density $\rho_{c,14}$, quark core radius $R_{Q}$, quark plus hadronic core
    radius $R_{Q+H}$, the isothermal core radius $R_{cr}$, the star radius $R$, and the masses of the quark core
    $M_{Q}$, the hadronic  core $M_{H}$,  the crust $M_{cr}$, and the
    total mass  $M$.  The density is in 
    units of $10^{14}$ g cm$^{-3}$, the radii in kilometers, and the masses in units of $M_{\odot}$~\cite{2011PhRvD..84f3015H}. 
  }    
		\begin{tabular}{ccccccccc} 
\\
\hline
\hline\\
	   $\rho_{c,14}$ & $R_{Q}$ & $R_{Q+H}$ & $R_{cr}$ & $R$
           &$M_{Q}$  & $M_{H}$ & $M_{cr}$ & $M$ \\
\\
\hline
\hline\\	  
	    5.1&-&11.47&13.39&13.53&-&1.03&0.07&1.10\\
	    8.2&-&12.57&13.55&13.57&-&1.81&0.04&1.85\\
\\
\hline
\\
	    10.8&0.68&12.54&13.49&13.5&0.001&1.866 & 0.0397 &1.906\\
	    11.8&3.41&12.40&13.31&13.32&0.093 &1.802 &0.0374&1.932\\
	    21.0&6.77&11.34&11.91&11.92&0.888 &1.143 & 0.0191&2.050\\
\\
	  \hline\hline\\
		\end{tabular}
	\label{table:1}  
\end{table}

The parameters of our models are listed in table \ref{table:1}.  The models
with low central densities ($\rho_c\le 3\rho_0)$ have masses
$M\le 1.85 M_{\odot}$ and are purely hadronic. The models with larger
central densities contain quark cores, with the size of the core
increasing with the central density.

\subsection{Neutrino emissivities in quark phases}
\label{sec:NeutrionEmissivities}

We consider first unpaired quark matter. Because quarks are
ultra-relativistic the Urca process works without any kinematical
restrictions (as opposed to baryonic matter). For two flavor quark
matter consisting of $u$ and $d$ flavors of quarks the main modes of
neutrino emission are given by
\bea
d \to u+e+\bar \nu, \qquad  u+e\to d + \nu, 
\eea 
where $e$ stands for electron, $\nu$ and $\bar \nu$ 
electron neutrino and antineutrino. The corresponding emissivity
(energy output per unit volume and time) was calculated to leading
order in the strong coupling constant $\alpha_S$; the  emissivity 
 {\it per quark color} is given by~\cite{1980PhRvL..44.1637I},
\begin{equation}\label{eq:iwamoto} 
\epsilon_{\beta} =\frac{914}{945}G^2\cos^2
\theta p_d p_up_e\alpha_sT^6 , 
\end{equation}
where $G$ is the weak coupling constant, $\theta$ the Cabibbo angle,
and $p_d$, $p_u$, and $p_e$ are the Fermi momenta of down
quarks, up quarks, and electrons.

The quark pairing modifies the temperature dependence of process
(\ref{eq:iwamoto}). If the quark condensate was a BCS-type
superconductor, the emission would have been suppressed linearly for
$T\simeq T_c$ and exponentially for $T\ll T_c$, where $T_c$ is the
critical temperature. However, in the gapless superconductors there is
an additional new scale $\delta\mu = (\mu_d-\mu_u)/2$, where
$\mu_{u,d}$ are the chemical potentials of light quarks.  Depending on
the relative ratio of the two scales in the problem, namely
$\delta\mu$ and $\Delta_0$ defined as the pairing gap for
$\delta\mu=0$ the suppression of emissivity is qualitatively
different. In terms of the parameter $\zeta = \Delta_0/\delta\mu$ one
finds that in the $\zeta > 1$ regime the Fermi surfaces of quarks are
gapped and the emissivity is suppressed as in the case of BCS
superconductors ~\cite{2006PhRvC..73d2801J}. In the opposite case
$\zeta < 1$ the Fermi surfaces have nodes and particles can be excited
around these nodes without energy cost needed to overcome the gap.  In
the case of the FF phase the shift in the chemical potential is
replaced by a more general function - the anti-symmetric in the flavor
part of the single particle spectrum of up and down quarks. As a
result the parameter $\delta \mu$ is replaced by
$ [\epsilon_d(\vec Q)-\epsilon_u(\vec Q)]/2$, where $\vec Q$ is the
total momentum of a Cooper pair and $\epsilon_{u/d}$ are the spectra
of the $u$ and $d$ quarks.

It is convenient to use below the generic parameterization of the
suppression factor of the quark Urca process given by~\cite{2006PhRvC..73d2801J} 
\bea \epsilon^{rg}_{\beta}(\zeta;T\le T_c) = 2f(\zeta)
\epsilon_{\beta}, \quad f(\zeta) =
\frac{1}{\exp[(\zeta-1)\frac{\delta\mu}{T}-1]+1}. \nonumber\\
\eea 
A time-independent constant $\zeta$ excludes the possibility of the
phase transition between the above mentioned phases.  Therefore,
following Ref. ~\cite{2013A&A...555L..10S}, we adopt {\it temperature
  dependent} (and, therefore, time-dependent) parameter $\zeta (T)$.
We use the following parameterization
\be \zeta(T) =
\zeta_i - \Delta\zeta ~g(T), 
\ee 
where $\zeta_i$ is the initial (pre-transient in the Cas A case) value,
$\Delta\zeta$ the constant change in this function, and the function
$g(T)$ describes the transition from the initial value $\zeta_i$ to
the asymptotic final value $\zeta_f = \zeta_i - \Delta\zeta$. The
transition is conveniently modeled by the following function \be g(T)
= \frac{1}{\exp\left(\frac{T-T^*}{w}\right)+1}, \ee which allows us to
control the temperature of transition, controlled by $T^*$, and the
smoothness of the transition, controlled by the width $w$.

For the blue quarks, which are not involved in the FF pairing, we 
assume that these are paired as in the BCS phase, as already
discussed in the previous section. In that case their emissivity in
the superconducting phase is very approximately related to the
emissivity in the normal phase 
\bea\label{eq:blue_suppress}
  \epsilon^{b}_{\beta}(T\le T_c)
  \simeq \epsilon_{\beta}^b (T > T_c) \,
 J_b\left(\frac{\Delta_b}{T}\right) \exp\left(-\frac{n\Delta_b}{T}\right) ,
\eea
where the function $J_b$ is some power-law function and $n$ is an
integer of order of unity. We assume $J_b = 1$ as it is of order unity
when $T \sim \Delta_b$ and is dominated by the exponential factor for
$T\ll\Delta_b$.  The pairing gap $\Delta_b$ will be treated as a fit
parameter below.

\subsection{Neutrino emissivities in hadronic phases}

Our treatment of the hadronic phases follows the standard picture of
slow cooling, which is supplemented by the pairing-breaking processes
in the superfluid phases.  Because of low proton fraction ($x_p\sim 0.05$)
in our models the dominant neutrino emission channel in the unpaired
phases of hadrons is the modified Urca process. Additional smaller
contribution from the modified bremsstrahlung process is also
included. We use emissivities derived for the free pion-exchange model
of strong interaction~\cite{1979ApJ...232..541F}, but reduce them by a
constant factor 5 to account for short-range repulsive component of
the nuclear force~\cite{1995MNRAS.273..596B,2002PhRvC..65f4007T}. The
effective mass corrections are included as well.

Below the critical temperatures ($T_c$) for neutrons and protons the
emissivities of these processes are suppressed, at asymptotically low
temperatures exponentially. In general the emissivity of
the modified Urca process in the superfluid phase can be  expressed
via its rate in the normal matter as follows
\bea 
\epsilon_{\beta\,{\rm mod}}
(T\le T_c) &=& \epsilon_{\beta\,{\rm mod}}(T > T_c) \nonumber\\
&\times&J_{\beta\,{\rm mod}}
\exp\left(-\frac{m\Delta_n(T)+k\Delta_p(T)}{T}\right),
\eea
where $m$ and $k$ are integers and $J_{\beta\,{\rm mod}}$ is a
power-law
function of $\Delta_p/T$ and $\Delta_n/T$, where $\Delta_{n,p}$ are
the pairing gaps of protons and neutrons.  The emissivities of the
nucleonic superconducting phases were derived in
Refs.~\cite{Gusakov2002,Gusakov2001} for both modified and direct Urca
as well as in Ref.~\cite{Sedrakian2005,Sedrakian2007} for the direct
Urca process using different methods. Below, for sake of simplicity,
we assume $m=k=1$ as well as $J_{\beta\,{\rm mod}}= 1.$ This
approximation captures approximately the low-temperature asymptotic
reduction of the emissivity, but is a rather crude approximation at
 $T\le T_c$.  More detailed expressions are not needed for the
description of cooling of hybrid stars, which is dominated by the
processes involving quarks.
The neutrino emission processes from Cooper PBF start to
contribute below respective $T_c$~\cite{1976ApJ...205..541F,1987PhLB..184..119S}.  The
neutral vector current processes are strongly suppressed by multiloop
contribution to the response function of $S$-wave paired
condensate~\cite{2006PhLB..638..114L,2007PhRvC..76e5805S,2008PhRvC..77f5808K,2012PhRvC..86b5803S}.
The axial-vector emission can be taken at one-loop level to a good
accuracy.  For $P$-wave paired neutrons the rates are unaffected by
the multiloop processes and we use the results of
Refs.~\cite{1999A&A...343..650Y,2011PhRvC..84d5501L}.  Given the
zero-temperature values of the nucleonic gaps $\Delta(0)$, their
finite-temperature values are well approximated by the asymptotic expressions 
given in Ref.~\cite{1959ZPhy..155..313M} (as well as fits to tables in 
that work \cite{HessDiploma}) 
\bea\label{gap_T} \frac{\Delta(T)}{\Delta(0)} =
\left\{\begin{array}{cc} 1- \sqrt{2 \gamma \tau} e^{-\pi/(\gamma\tau)}
    & 0\le
      \tau\le 0.5, \\
         \sqrt{3.016(1-\tau) -2.4(1-\tau)^2} & 0.5< \tau\le 1
                                               ,\end{array}\right.
\eea
where $\Delta(0)$ is the pairing gap in the zero-temperature limit,
 $\tau = T/T_c$ is the temperature in units of the critical
temperature and $\gamma = 1.781$.  The formulae \eqref{gap_T}
reproduce the solution of the BCS gap equation with zero-range
interaction within a percent accuracy.
Crustal neutrino emission contributes to the cooling of the star in
the final stages of neutrino cooling era via the electron neutrino
bremsstrahlung emission on nuclei in the
crusts~\cite{1969PhRv..180.1227F}.  Ions in the crust may form a fluid
or a solid.  The abundances of impurities, if high, could lead to
additional the neutrino emission. In the fluid or impurity dominated
crust the emissivity scales as $T^6$ and we adopt this option for
temperatures relevant for that particular cooling phase
($T \le 10^9 K$).  If crystalline lattice is formed at the temperatures
of interest, then the emissivities would scale approximately as $T^7$
(classical crystal) or as $T^8$ (quantum crystal) and the
bremsstrahlung emissivity is parametrically
suppressed~\cite{1999A&A...343.1009K}.

\subsection{Specific heat }

The temperature dependence of the specific heat of normal 
Fer\-mi-liquids (both relativistic and non-relativistic) 
is given by the linear law $c_V = a T$, where the coefficient depends on 
the abundance of given species (which is the same as the Fermi 
momentum).  We apply this formula to the leptonic component of the 
star and unpaired hadrons and quarks. 

As well known, in ordinary BCS superconductors the specific heat as a
function of temperature experiences a jump at the corresponding
critical temperature and then decays, at low temperatures
exponentially. For $S$-wave superconductors one can apply the a
asymptotic  expressions given in Ref. ~\cite{1959ZPhy..155..313M} (as well
as fits to tables in that work \cite{HessDiploma}) to model the ratio
of the specific heats in the superconducting $c_S$ and normal $c_N$
phases
 \bea\label{cv_blue} \frac{c_S (T)}{c_N (T_c)} =
\left\{\begin{array}{cc} (12\pi/\gamma) 
    (2\gamma\tau)^{-3/2}e^{-\pi/(\gamma\tau)}& 0\le \tau\le 0.3,    \\
-0.24422+0.255292\tau\\ 
  +2.43077\tau^2 & 0.3< \tau\le 
    1, \end{array}\right.  
\eea 
where $T_c$ is the critical temperature of phase transition. 
Equation~(\ref{cv_blue}) was applied uniformly to neutron, proton, and 
blue-quark condensates. See Ref.~\cite{1994ARep...38..247L} for
possible effects of $P$-wave superfluidity of neutrons in the core on
their specific heat.

To find the  specific heat of the  red-green condensate we need 
an expression for the critical temperature as a function of the 
mismatch $\delta\mu$. We use the following formula 
 \be 
T_c(\zeta) \simeq T_{c0}\sqrt{1-\frac{4\mu}{3\Delta_0} \delta 
  (\zeta)}, 
\ee 
where $\mu = (\mu_d+\mu_u)/2$, $\Delta_0 = \Delta(\zeta=0)$,
$T_{c0} = T_c(\zeta=0)$, and $\delta (\zeta) = (n_d-n_u)/(n_d+n_u)$. 

As in the case of emissivities, the availability of gapless fermions 
in the case $\zeta\le 1$ changes the BCS behaviour of the specific
heat as well.  To model this regime we apply, in analogy to
emissivities, the following formula
\bea 
\frac{c_S^{rg}(\zeta;T\le T_c)}{c_N^{rg}} &=& f(\zeta),
\eea 
where $c_N^{rg}$ is the specific heat of red-green unpaired quarks,
taken as that for noninteracting quarks and 
$c_S^{rg}$ is the specific heat of pair-correlated quarks. 

\section{Results of numerical simulations}
\label{sec:NumRes}

\subsection{General considerations}

The thermal evolution code employed in our study uses the isothermal
core approximation which is valid for timescales that are larger that
those which are required to dissolve temperature gradients by thermal
conduction. In the hadronic core the thermal conductivity is dominated
by the electron transport if baryons are superfluid
\cite{1979ApJ...230..847F} and baryons if these are non-superfluid \cite{2007PhRvD..75j3004S}.
 The
characteristic times-scales for thermal relaxation are of the order of
$t\sim 100$~y.  In the 2SC phase likewise the transport is dominated
by electrons and blue-quarks~\cite{2014PhRvC..90e5205A}. 
the early stage of thermal evolution of 2SC phase and the time-scale
of its thermal relaxation remains to be studied; here we assue that
the entire (quark plus hadronic) core is thermally relaxed at the start
of simulation.  Note that for reasonable initial temperatures the
cooling tracks exit the non-isothermal phase and settle at a
temperature predicted by the balance of the dominant neutrino emission
mechanism
and the specific heat of the core independent of the earlier evolution
phase.  

The isothermal core is defined by the transition density
$\rho_{\rm tr}=10^{10}$ g cm$^{-3}$. At lower densities the envelope
maintains temperature gradients throughout the complete evolution.
The physics of thermal transport can be encoded in a relation between
the interior and surface temperatures. The models of envelopes predict
the scaling $T_s^4 = g_s h(T)$, where $T_s$ is the surface temperature 
$T$ is the the isothermal core temperature, 
$g_s$ is the surface gravity,
and $h$ is a function which depends on $T$, the opacity of crustal
material, and its EoS. This can be written
as~\cite{1983ApJ...272..286G,1997A&A...323..415P}
\bea 
T_{s6} = (\alpha  T_9)^{\beta} g_s^{1/4} 
\eea 
where $T_9$ the isothermal core temperature in units of $10^9$ K,
$T_{s6}$ is the surface temperature in units of $10^6$ K, $\alpha$ and
$\beta$ are constants that depend on the composition of the star's
atmosphere, $g_s$ is the surface gravity in units of $10^{14}$ cm
s$^{-2}$. We use $\beta= 0.55$, which lies between the values for the
purely-iron ($\beta = 0.5$) and the fully accreted ($\beta = 0.61$)
envelope and $\alpha = 18.1$, which is appropriate for an accreted
envelope~\cite{1997A&A...323..415P}.  Our fit results are sensitive to
the $\beta$ parameter, but not to the $\alpha$ parameter.  With these
ingredients at hand we have numerically integrated the thermal
evolution equation as described in Ref.~\cite{2011PhRvD..84f3015H}.

The numerical simulations were carried out for all models listed
in Table~\ref{table:1}, but we concentrate below mostly on the
representative model with $M/M_{\odot} = 1.93$. Three parameters were
varied: (a) the transition temperature $T^*$ from the 2SC to the
crystalline phase; (b) the blue quark gap $\Delta_b$; and (c)
the width of the transition $w$. The remaining parameters which model
the temperature (time) dependent part of the $\zeta$ function were held
fixed at values $\zeta_i = 1.1$ and $\Delta\zeta = 0.2.$ The constant
zero-temperature value of the red-green gap is fixed at the value
$\Delta_{rg}=60$~MeV.
\begin{figure}[t]
\begin{center}
\includegraphics[width=\linewidth,height=7cm]{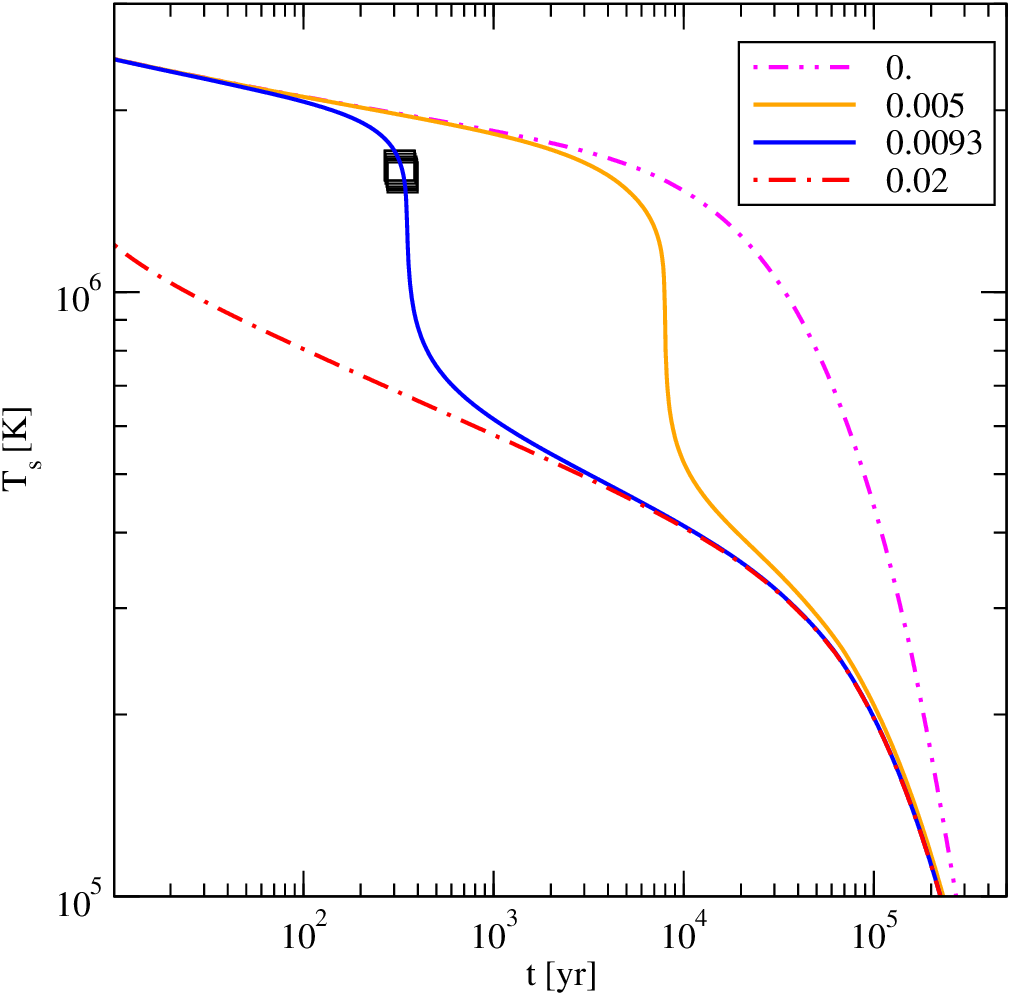}
\caption{ Dependence of the surface temperature of a CSs 
  given in K on time in year. The lines correspond to different phase 
  transition temperature $T^*$ for fixed values of the 
  width $w = 0.2$ and the blue quark pairing gap $\Delta_b= 0.15$
  (both in MeV). The labels correspond to values of $T^*$ in MeV.  The 
Cas A data is shown by squares.}
\label{epja:fig3}
\end{center}
\end{figure}
Figure \ref{epja:fig3} shows the dependence of the (redshifted)
surface temperature on time for $M/M_{\odot} = 1.93$ stellar model and
for different values of the $T^*$ (as indicated in the figure) with
$w$ and $\Delta_b$ fixed. In the limit $T^*\to 0$ the quark core does
not influence the cooling because 
the neutrino emission from the red-green and blue
condensates is suppressed for $T\ll \Delta_{rg}$ and $T\ll
\Delta_b$.
For high values of $T^*$ (0.2 MeV in the figure) the transition to
the FF phase occurs early in the evolution of the star, therefore
enhanced neutrino emission cools the star rapidly below the
value observed for the CS in Cas A. It is further seen that by tuning
the phase-transition temperature to the value $T^*=0.009385 $ MeV the
``drop'' in the temperature of the model can be adjust to the observed
temperature and age of the CS in Cas A. Thus, if the current
interpretation is correct, then the CS is a massive compact star 
(so it features a quark core) and it undergoes currently a phase
transition from the 2SC to the FF phase (or some other gapless
phase). Note that the deduced phase transition temperature is of the
order of 0.01 MeV and is small compared to the scale set by the
$\Delta_{rg}$. This could be the feature of the equilibrium phase
diagram of the superconducting quark matter or alternatively may
reflect the non-equilibrium aspect of the phase transition, i. e.,
long-lived metastable 2SC phase (in parallel to super-heating or
super-cooling of superconducting states, know from the physics of
ordinary superconductors). We conclude that with a proper choice of
the value of the transition temperature we can account for the
observed ``drop'' in the temperature.
\begin{figure}[t]
\begin{center}
\includegraphics[width=\linewidth,height=7cm]{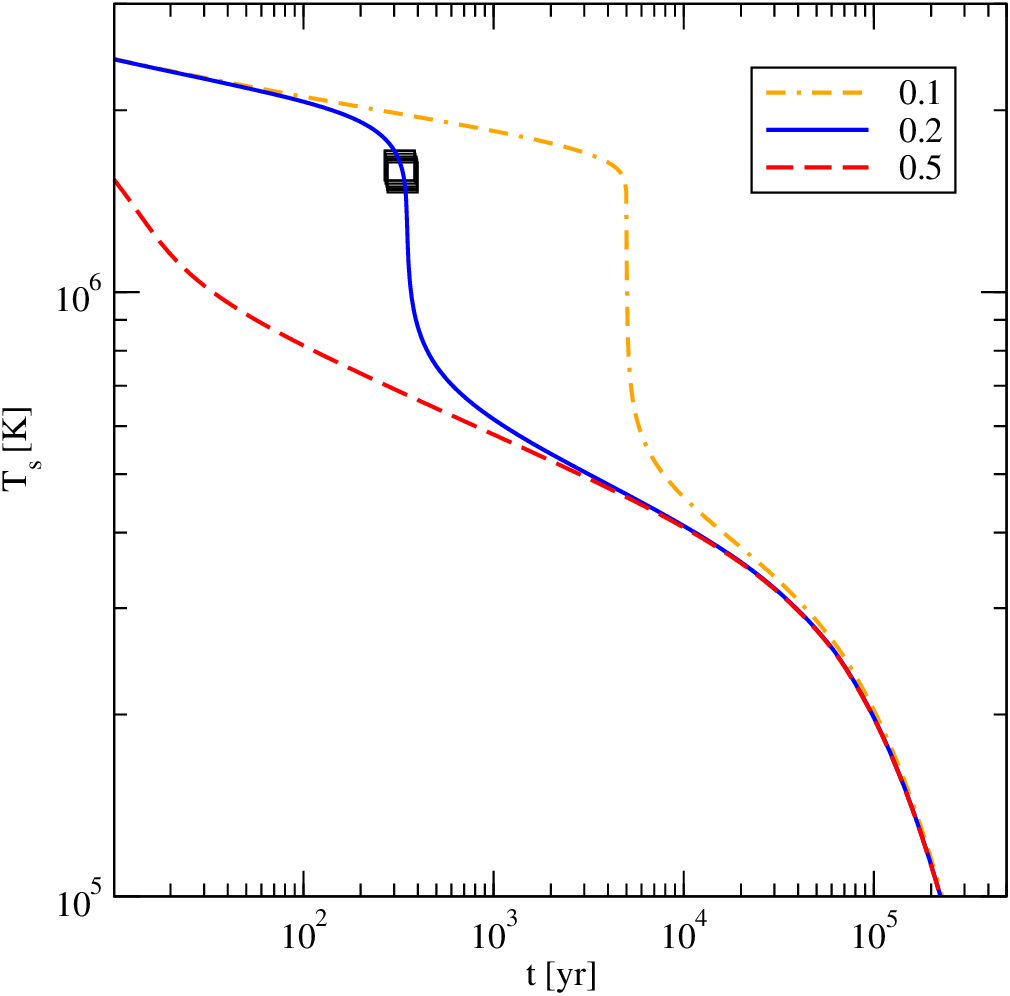}
\caption{Same as in fig.~\ref{epja:fig3} but for different values of the width $w$
 for fixed $T^* = 0.009385$ and  $\Delta_b=0.15$ MeV.
The labels correspond to values of $w$ in MeV. 
}
\label{epja:fig4}
\end{center}
\end{figure}

Next we fix the parameters $T^* $ and $\Delta_b$ and vary the width of
the transition $w$, which controls the ``smoothness'' of the
transition (fig.~\ref{epja:fig4}). For small values of $w$ (0.1 in the
figure) the transient is steeper, as expected, but is also delayed
(i. e. the drop is shifted away from the Cas A location to later
times). Inversely, large $w$ make the transient much smoother and also
the start of the phase transition is shifted to the left, i. e., to
earlier times. Thus, small (by a factor of 2) variation of the
parameters $w$ and $T^*$ can cause substantial shifts in the cooling
curves, i. e., the overall fits to the Cas A data are sensitive to the
values of these parameters.
\begin{figure}[t]
\begin{center}
\includegraphics[width=\linewidth,height=7cm]{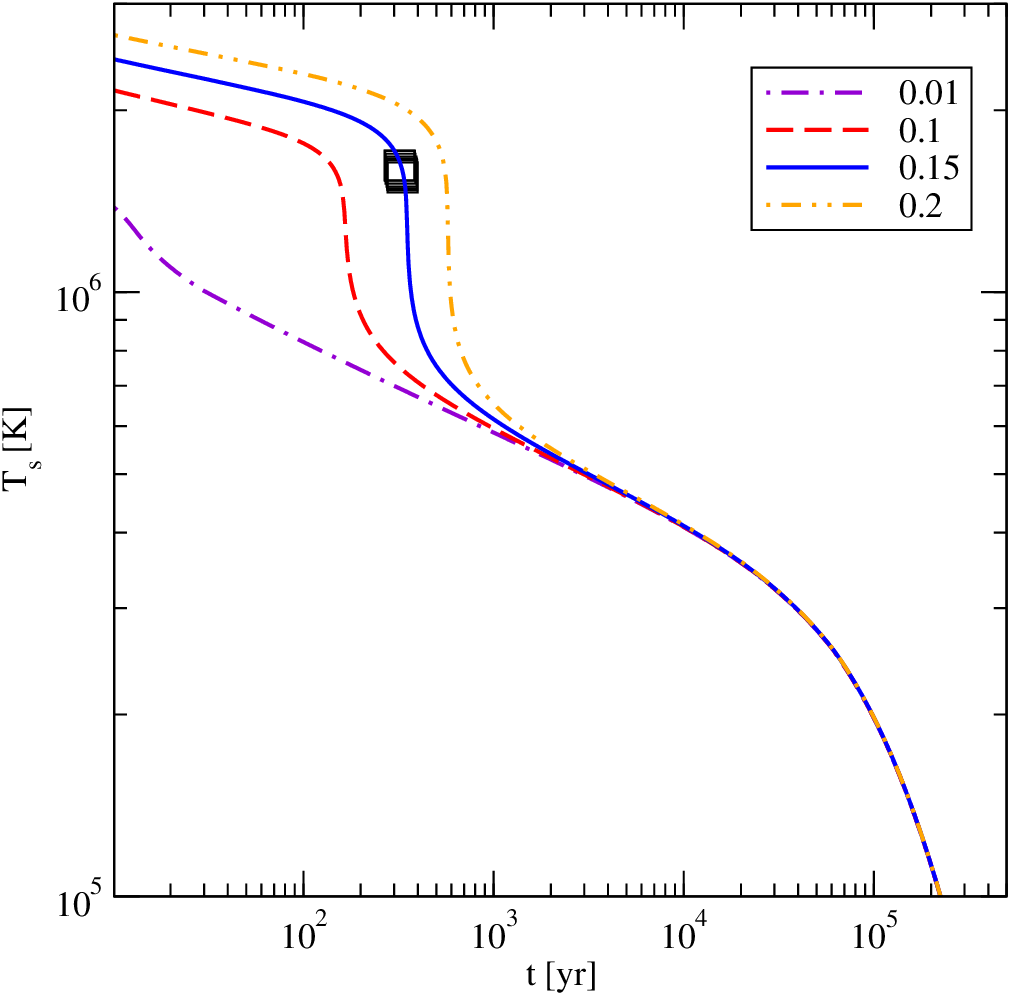}
\caption{Same as in fig.~\ref{epja:fig3} but for different values of the 
  blue quark gap $\Delta_b$ for fixed $T^* = 0.009385$ and $w=0.2$
  MeV. The labels correspond to values of $\Delta_b$ in MeV. 
}
\label{epja:fig5}
\end{center}
\end{figure}
Finally we fix the parameters $T^* $ and $w$ and vary the value 
of the gap of blue quarks $\Delta_b$ (fig.~\ref{epja:fig5}). Increasing the
magnitude of $\Delta_b$ shits the transient to later times, but also
the pre-transient temperatures increase because of the
exponential suppression factor in eq.~\eqref{eq:blue_suppress} is less
effective. Inversely, decreasing the value of $\Delta_b$ shifts the
transient to earlier times and decreases the transient temperature.
This behaviour is consistent with the fact that the cooling for heavy
models is dominated by the red-green and blue quark components. In the
2SC phase the cooling is controlled by the Urca process on blue
quarks, which is  suppressed for late times and low temperatures, see
Eq.~\eqref{eq:blue_suppress}. Unless the value of $\Delta_b$ is too
small the blue quarks dominate throughout the neutrino cooling era. At
later stages $t\sim 10^5$ y the photon emission becomes the main
cooling mechanism (this is indicated by the change of the slope of the
cooling curve). During and after the transient, 
the gapless phase emits neutrinos on
red-green quarks; their luminosity is larger by  factor two - the number of
involved flavors -  compared to blue quark luminosity. In this
case again the transition is to the photon cooling at $t\sim 10^5$ y. 
An important conclusion is that for massive CS the neutrino cooling is
completely dominated by quarks and the details of the modeling of
hadronic neutrino cooling are irrelevant. This implies, obviously, that
the quality of the fits to the entire population of thermally emitting
pulsars is not correlated to our fits to the Cas A data, as this
population needs to include purely hadronic neutron stars with $M\le
1.8 M_{\odot}$, which will cool much slowly.

\begin{figure}[t]
\begin{center}
\includegraphics[width=\linewidth,height=7cm]{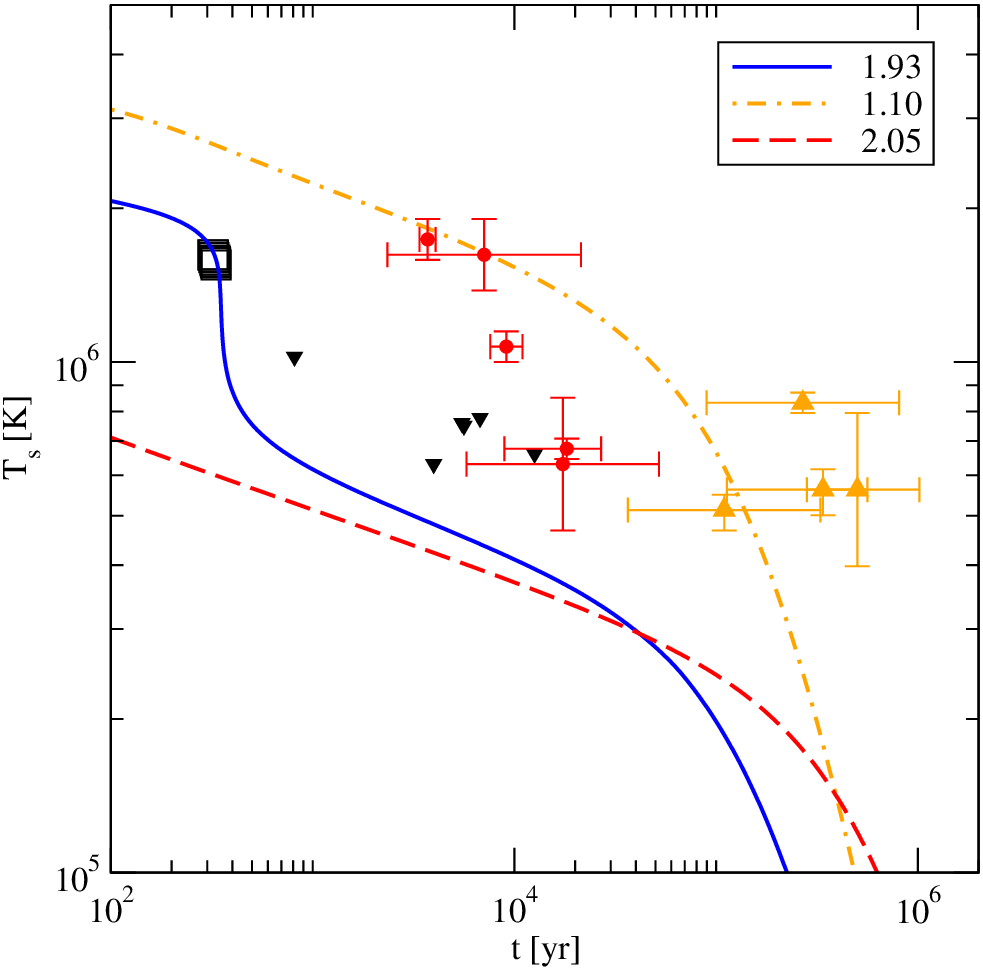}
\caption{Same as in fig.~\ref{epja:fig3}  for fixed $T^* = 0.009385$
  but for different masses of CS. The labels correspond to masses of
  the stars in units $M/M_{\odot}$. The references in the data are
  given in  fig. 3 of Ref.~\cite{2011PhRvD..84f3015H}.
}
\label{epja:fig6}
\end{center}
\end{figure}
We show in fig.~\ref{epja:fig6} the variation of cooling curves with
the mass of the CS holding the ``best'' values of the parameters
fitting Cas A fixed.  It is seen that there is a substantial variation
in the cooling speed of CSs with the mass; the heavy stars are cooling
much faster than the light stars, for the neutrino emissivity of the
hadronic phases is less effective that those of the quark phases. Note
also the inversion of the temperatures in the photon cooling era: here
the heavier $M/M_{\odot } = 2.05$ star is hotter than the lighter
$M/M_{\odot } = 1.93$ star.  The band lying between the cooling curves
of extrem mass object, i.e., $1.1\le M/M_{\odot } \le 2.05$ in the
$T$-$t$ plane can be covered by changing the central density (and therefore
the mass) of the model and the experimental data can naturally be
accommodated. (The few outliers, which are hot at late stages of
evolution $t\sim 10^6$ y, may reflect different surface composition,
than assumed in our simulations, e.g., composition containing
hydrogen). Thus, we conclude that the spread in the temperatures
of thermally emitting neutron stars can be attributed to the different
masses of these objects: the lighter ones not containing quark matter
cool more slowly than the heavier stars having quark cores. Of course,
variations in other factors, such as the strength of the magnetic field,
the surface composition, etc can also contribute to the dispersion.
\begin{figure}[t]
\begin{center}
\includegraphics[width=\linewidth,height=8cm]{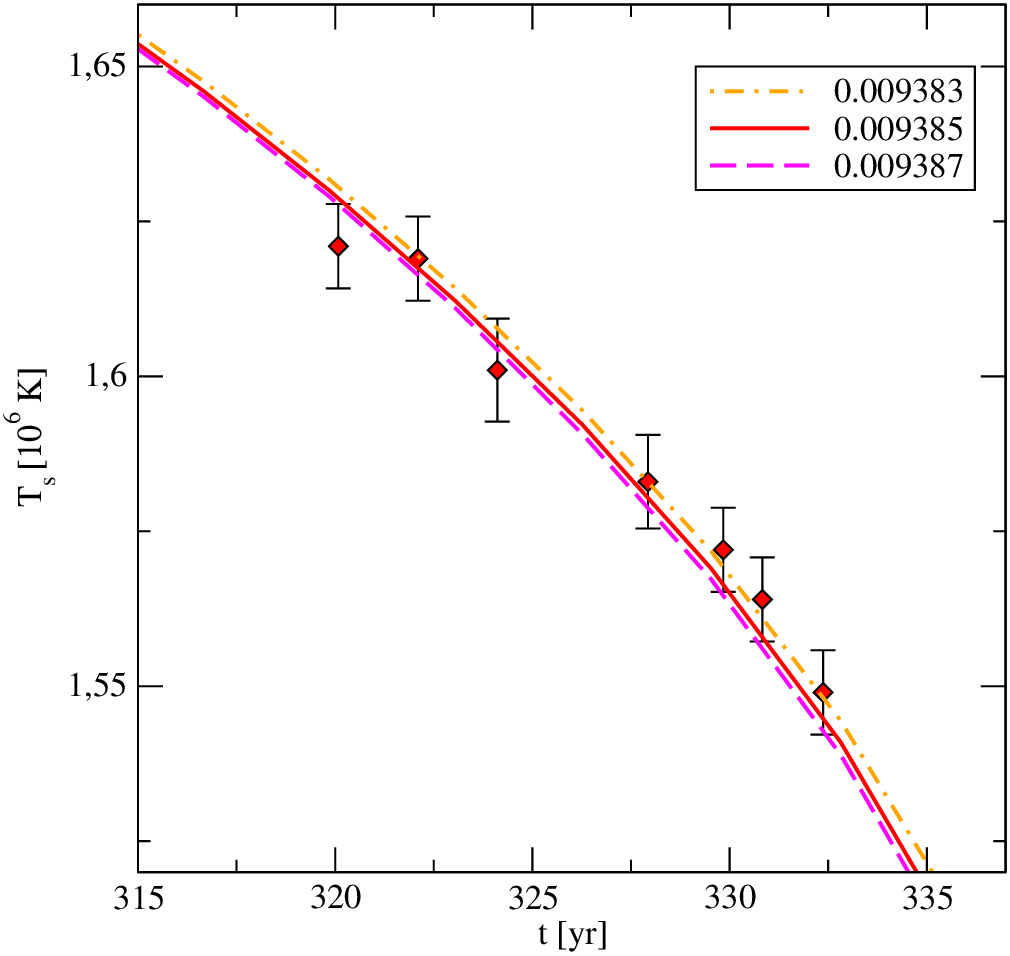}
\caption{Dependence of the surface in units of $10^6$ K temperature on
  time for several temperatures of the  phase transition  $T^*$
 at fixed $w = 0.2$ and $\Delta_b = 0.15$ (both in MeV). The dots with
 error bars show the Cas A data taken from~\cite{2013ApJ...777...22E}.}

\label{epja:fig7}
\end{center}
\end{figure}
\begin{figure}[hbt]
\begin{center}
\includegraphics[width=\linewidth,height=8cm]{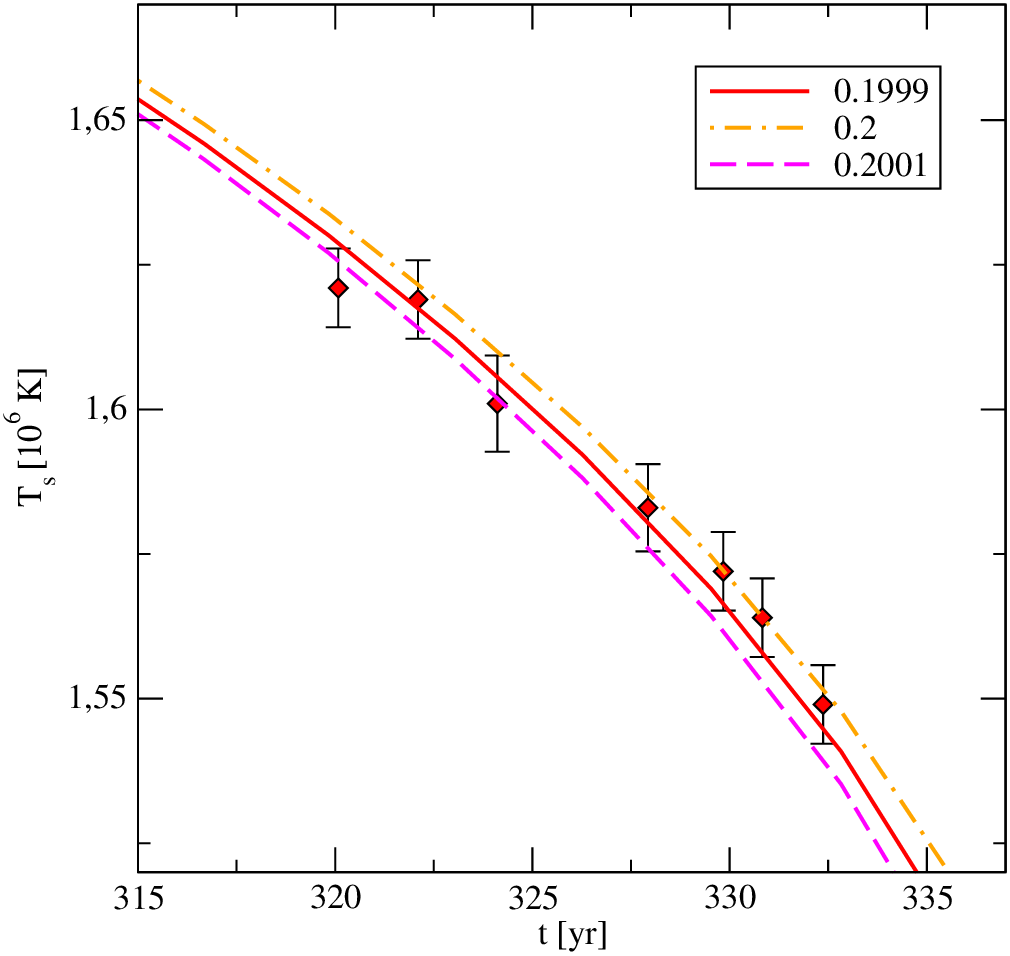}
\caption{Same as in \ref{epja:fig7}, but for a range of values of $w$
  indicated in the plot and for fixed values of  $T^* = 0.009385$ and
  $\Delta_b = 0.15$. All parameter values are in MeV.}
\label{epja:fig8}
\end{center}
\end{figure}
\begin{figure}[thb]
\begin{center}
\includegraphics[width=\linewidth,height=8cm]{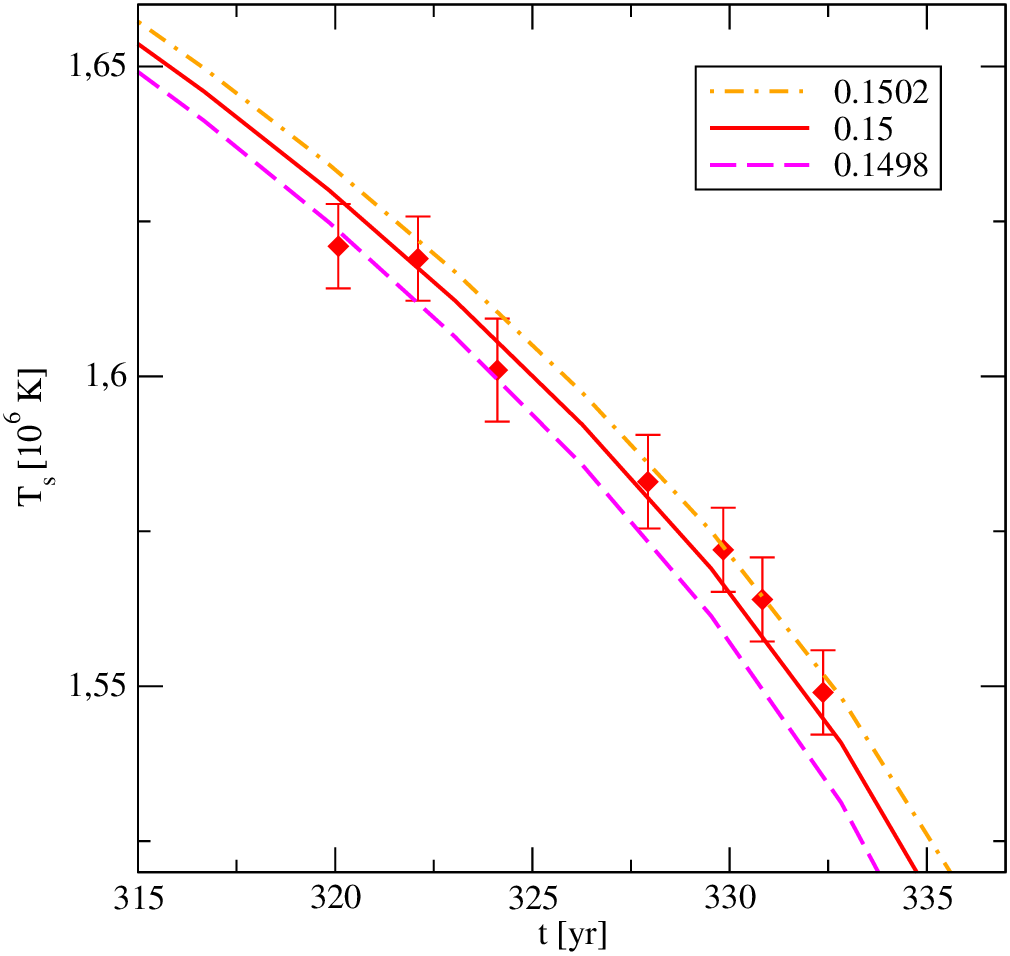}
\caption{Same as in \ref{epja:fig7}, but for a range of values of $\Delta_b$
  indicated in the plot and for fixed values of  $T^* = 0.009385$ and
  $w = 0.2$. All parameter values are in MeV.}
\label{epja:fig9}
\end{center}
\end{figure}
\subsection{Fits to the Cas A data}
The Cas A data spanning the decade 2003-13 have been used for the
fits. It is assumed that the star is already in the transient stage,
as there is no evidence of the entry into this phase from a slower
cooling stage, as well as no sign of the exit from the fast cooling
mode. The parameters were adjusted to obtain the location of the Cas A
data in the $T$-$t$ diagram as well as the slope of the transient. The
best fit is obtained with the parameter values $T^*=0.009385$,
$w = 0.2$ and $\Delta_b=0.15$ (all quantities in MeV). The sensitivity
of fits was tested against variation of one of them at fixed values
of the other two.  Figure~\ref{epja:fig7} shows the effect of
variations of $T^*$ on the transient behaviour. We quantify the
sensitivity of the fits by defining the relative deviation as
$\delta T^*/T^*$, where $\delta T^*$ is the range of values of $T^*$
compatible with the data; we find that
$\delta T^*/T^* \simeq 10^{-3}$.  Figure~\ref{epja:fig8} shows the
effect of variation of the width $w$ at fixed values of other two
parameters. In this case we find that
$\delta w /w = 1.5\times 10^{-3}$, i.e., the relative deviation is 
in the same range as for the transition temperature $T^*$.
Figure~\ref{epja:fig9} shows the variation in the value of the blue
quark gap $\Delta_b$; in this case we find that
$\delta\Delta_b/\Delta_b\simeq 3.3\times 10^{-3}$.

Thus we conclude that the fitted values are quite sensitive to the
precise values of parameters and small deviations at the level of
$10^{-3}$ can spoil the fit to the Cas A data.  Extrapolating from the
present fits one would predict that  the CS in Cas A would continue
the fast cooling up to temperatures somewhat below $10^6$ K. Clearly,
an observation of an exit from the transient would provide further
strong constraints on any model of the behaviour of the CS in Cas A and
current fits should by updated to account for that effect.

\section{Summary and conclusions}
\label{sec:Conclusion}

The observation of two-solar mass pulsars in binary systems is a
strong evidence that the EoS of dense matter must be stiff and that
the densities in the centers of compact stars can be large enough
(several times the saturation density) so that the threshold density
for transition to quark matter can be reached.  This motivates the
studies of the dense phases of quark matter under the conditions
expected in compact stars (charge and color neutrality,
near-equilibrium with respect to weak processes). We have concentrated
here on CSs that have sharp interface between the hadronic and quark
matter phases, in which case the massive members of the sequences
($M> 1.85 M_{\odot}$) contain progressively larger amount of quark
matter whereas the lighter members are purely hadronic. We have
evolved these models in time and followed the changes in the core and
surface temperature under assumption that the core is isothermal,
which is a valid approximation, except for the first 100 y following
the birth of the star.

The generic phase diagram of imbalanced fermionic systems, which also
includes the flavor asymmetric pairing among quarks, implies that at
high temperatures the pairing is in the BCS type phase, i.e., both
Fermi surfaces of quarks are gapped despite of the mismatch between
the Fermi surfaces of $u$ and $d$ flavors. At lower temperatures a
transition from the BCS to a generic gapless phase must take
place. This gapless phase will have some type of spatial modulation
and we considered the FF type simple realization of such phase. The
neutrino emissivity of such phase must be much larger than the
emissivity of the 2SC phase because of the existence of gapless
excitations. We have modelled the emissivities of both phase in terms
of a simple parameterization, which however takes into account the
fact that as the temperature is lowered there is a phase transition
from the 2SC to FF phase.

The rapid cooling of the CS in Cas A can be accounted for via the
phase transition described above; we stress again that the phase
transition takes place within the phase diagram of QCD and is the
consequence of the ordering of various superconducting phases in the
temperature, density, and isospin spaces. Such ordering was observed
in numerous studies of other imbalance systems such as the hadronic
superfluids or ultra-cold atomic condensates.

The Cas A data can be fitted by varying the phase transition
temperature $T^*$ at fixed value of the remaining physical parameters
- the width of the phase transition (or, equivalently, its duration)
and the gap in the spectrum of BCS-paired blue quarks. It turns out
that the relative accuracy of order $10^{-3}$ is required to fit the
data, i. e., larger deviation will spoilt the fit. We also tested the
sensitivity of the other two parameters to the variations when the
remaining two parameters are fixed. The relative accuracy is again in
the range $10^{-3}$. Assuming that the present model of cooling of CS
in Cas A is correct, we immediately conclude that only the massive members
of CS sequences that contain (color-superconducting) quark matter can
undergo transients of this type. This is the 
main difference of our model to the hadronic alternatives which work
also for low (or canonical 1.4 $M_{\odot}$) stars. It is tempting to
think that the various models of cooling of CS in Cas A can be
distinguished through the measurements of the mass or other integral
parameters of the CS in the future.  

In addition to the Cas A behaviour we also studied the effect of
variations of the mass of the CSs on their cooling behaviour. First, we
find that larger (close to the maximum) mass stars will not show the
same transient as the assumed model with 1.93 $M_{\odot}$ mass; rather
they will cool to lower temperatures much faster. As our code does
not cover the very early evolution of CSs, we cannot resolve the
transient that takes place very early in the evolution of a highly
massive CS.  Secondly, we find that low mass stars $M\sim 1.1\,  M_{\odot}$ are
much hotter than their heavy counterparts, so that the band covered by
the stars in the range $1.1\le M/M_{\odot}\le 2$ can account for the
measured surface temperatures of thermally emitting CS.

There are a number of issues that need further attention: (a) The
appearance of strangeness in form of hyperons in the hadronic phase
and strange quarks in the quark phases can substantially affect
cooling. In particular strange quarks may even change the structure of
the phase diagram of color superconducting matter at high density by
inducing new phases and new channels for neutrino emission see, for
example, Ref. ~\cite{2005PhRvD..71k4011A}; (b) heating in the core and crust
of the star can become important at later stages of
evolution; it does not affect the
cooling at the age of CS in Cas A (if the CS is not a magnetar); (c)
the density dependence of pairing gaps of red-green and blue condensates
can be a factor. The density of states at the Fermi surface increases
as 1/3 power of density, whereas the strong coupling constant
decreases logarithmically.  Therefore, one can envision a situation
where the red-green condensate has $\zeta <1$ only in part of the
core. This will reduce the effective volume of the quark phase
contributing to the fast cooling; (d) the superconducting quark phases
may gradually appear as a result of deceleration of the star, i.e.,
the CS can experience a rotation induced phase transition to a color
superconducting phase~\cite{2013A&A...559A.118A}. This transition 
will occur in the densest region (center of the CS) in parallel with 
deconfinement of quarks. 

\section*{Acknowledgment}

This work ws supported by the Deutsche Forschungsgemeinschaft (Grant
No. SE 1836/3-1) and by the New\-Comp\-Star COST Action MP1304.

\bibliographystyle{}

\end{document}